\documentclass[usenatbib]{mn2e}
\usepackage{epsfig}
\usepackage{color}
\usepackage{amsmath,amssymb}
\newcommand{\hMsol}{{\,h^{-1}\rm M}_\odot}
\newcommand{\hMpc}{{\,h^{-1}\rm Mpc}}
\newcommand{\hkpc}{{\,h^{-1}\rm kpc}}
\newcommand{\kms}{{\,\rm km\,s^{-1}}}
\newcommand{\Gyr}{{\,\rm Gyr}}
\newcommand{\G}{{\,\rm G}}
\newcommand{\Myr}{{\,\rm Myr}}
\newcommand{\yr}{{\,\rm yr}}

\title[Oscillatory relaxation of a merging galaxy cluster]
{Oscillatory relaxation of a merging galaxy cluster}
\author[Faltenbacher et al.]
{\parbox[t]\textwidth{Andreas Faltenbacher$^1$, Stefan Gottl\"ober$^2$
and William G. Mathews$^1$} 
\vspace*{6pt} \\
$^1$UCO/Lick Observatory,
University of California at Santa Cruz, 
1156 High Street, Santa Cruz, CA 95064, USA
\\
$^2$Astrophysikalisches Institut Potsdam,
An der Sternwarte 16, 14482 Potsdam, Germany
}
\date{\today}
\begin{document}
\maketitle
\begin{abstract}
Within the cosmic framework clusters of galaxies are relatively young
objects. Many of them have recently experienced major mergers. Here we
investigate an equal mass merging event at $z\approx0.6$ 
resulting in a dark matter haloe of $\sim2.2\times10^{14}\hMsol$ at
$z=0$. The merging process is covered by 270 outputs of a high
resolution cosmological N-body simulation performed with the ART
(adaptive refinement tree) code. Some $2\Gyr$s elapse between the first
peri-centre passage of the progenitor cores and their final
coalescence. During that phase the cores experience six peri-centre
passages with minimal distances declining from $\sim30$ to
$\sim2\hkpc$. The time intervals between the peri-centre passages
continuously decrease from $9$ to $1\times10^8\yr$s. 
We follow the mean density, the velocity dispersion and the entropy
of the two progenitors within a set of fixed proper radii (25, 50,
100, 250, 500, $1000\hkpc$). During the peri-centre passages we find
sharp peaks of the mean densities within these radii, which exceed
the sum of the corresponding progenitor densities. In addition to the
intermixing of the merging haloes, the densities increase due to
contraction caused by the momentary deepening of the potential
well. The velocity dispersions also peak during peri-centre
passages. Within the fixed proper radii the entropy of the most
massive progenitor after the merger settles close to its pre-merger
values. At the end of the oscillatory relaxation phase the material
originating from the less concentrated of the two equal mass
progenitors is deposited at larger radii and shows a slightly more
radially anisotropic velocity dispersion compared to the material
coming from the more concentrated progenitor. Every peri-centre
passage is accompanied by a substantial drop of the central potential
well. We briefly discuss the possibility that AGN outbursts are
triggered by the periodically changing potential.
\end{abstract}
\begin{keywords}
cosmology:theory  -- galaxies:clusters:general --  methods:numerical
\end{keywords}
\section{Introduction}
The current cosmological model (e.g.~\citealt{Spergel06}) assumes the
gravitating matter to be composed of $~\sim 15$ per cent baryonic and
$~\sim 85$ per cent dark matter (DM). The former interacts
electromagnetically, the latter is not directly visible. By comparing
lensing observations with X-ray maps of the merging cluster 1E0657-558
\cite{Clowe06} find strong evidence for unseen matter concentrations
that are more massive than the baryonic component. These observations
emphasise the complexity of the interactions between dark and baryonic
matter during mayor merging events. Based on hydrodynamic simulations
of  head-on mergers between model galaxy clusters \cite{Pearce94} find
a damping of the core oscillations by the gaseous component. The
situation may become even more complex if the collision of gas-rich
systems can trigger star formation and outflows from the central
active galactic nucleus (AGN). Multi wave band observations by
\cite{Conselice01} reveal that this may actually the case for the
central galaxy NGC 1275 in the Perseus Cluster. 
 
Using high resolution gas dynamical simulations \cite{Nagai06}
demonstrate that the X-ray observations of unrelaxed clusters cannot
accurately measure the density and temperature gradients in the hot
baryonic intra cluster medium (ICM). In addition 
some potentially relevant physical processes, such as AGN bubbles, 
magnetic fields, and cosmic rays, are not well understood yet and
are usually excluded from such simulations. Here we do not aim to
model the behaviour of the gas directly. Our analysis is based on a
pure N-body simulation. However, since the collisionless component
dominates the gravitational potential, we hope to gain some   
complementary insight into the violent processes during a mayor merger
of galaxy clusters. The merging event is covered by a set of 270
simulation outputs which allows us to precisely measure even highly
variable features during the violent relaxation phase. Our goal is
twofold: (1) we provide a precise time table of events during the core
merging phase. The dynamics may be slightly altered by 
the presence of collisional material, however the general behaviour
is not expected to change profoundly. (2) we investigate the evolution
of the central potential to achieve some basic understanding of the
gravitational forces which the baryons are exposed to. Ultimately we 
wish to uncover interactions between the DM and the baryonic
components which result in observable phenomena. 

The paper is organised as follows. \S~\ref{sec:simda} introduces the
simulation and data preparation. In \S~\ref{sec:res} we present the
evolution of the density, the velocity dispersion and the entropy of
the DM within spheres of fixed proper radii. Additionally, the
evolution of the central gravitational potential is
discussed. Finally, \S~\ref{sec:sum} provides a summary of the main
results and a discussion focusing on the potential interplay between 
the dark and the baryonic component.
\section{Simulation and data preparation}
\label{sec:simda}
The N-body simulation of a cluster-sized DM haloe has been 
performed with the adaptive mesh refinement code {\sc ART}
\citep{Kravtsov97}. The present analysis is based on $\sim270$
snapshots between the redshifts $z=1.2$ and $z=0$. The cluster haloe
has been selected from an initial low-resolution simulation, 
$128^3$ particles within an $80\hMpc$ cube in a $\Lambda$CDM universe 
( $\Omega_m = 0.3$; $\Omega_\Lambda = 0.7$; $h = 0.7$; $\sigma_8 =
0.9$). Higher mass resolution has been achieved by means of the
multiple-mass technique \citep{Klypin01} applied to the particles
within a Lagrangian volume of approximately twice the virial radius
($2\times1.23\hMpc$) of the cluster at $z=0$. 
This technique results in a mass resolution of $3.2 \times
10^8\hMsol$ which is equivalent to an effective resolution of $512^3$ 
particles within the $80\hMpc$ box. The minimum cell size allowed in
the {\sc ART} runs was set to $1.2\hkpc$. The simulation started at
$z=50$ and the virial mass of the cluster at $z=0$ has is
$\sim2.2\times10^{14}\hMsol$ corresponding to $\sim700000$ particles 
within the virial radius. The haloe analysed here is the most massive
cluster in the  sample discussed by \cite{Ascasibar04}.

\begin{table}
\begin{center}
\begin{tabular}{crrr}
\hline$ll$&1&2&3\\\hline
0.20 & 338548  &&         41580\\
0.17 & 153916  & 140898 & 31656\\
0.10 &  71595  &  60984 & 15949\\
0.07 &  31077  &  31006 &  9733\\
0.02 &  3489   &   1312 &   326\\
\end{tabular}
\end{center}
\caption{\label{tab:memb}
Most massive FoF groups for linking lengths of $ll = 0.20$, 0.17,
0.10, 0.07, 0.02 times the mean particle distance at a redshift of
$z=1.2$. The three columns indicate the membership of
the (sub)structures to three spatially separated haloes. The 
omitted entry for the 0.20 groups means that the largest 0.20 group
includes the two largest 0.17 groups below.}
\end{table}
At $z\sim0.6$ two progenitors of similar mass merge. They are by far
the largest units assembled by the cluster up to the present time. We
wish to identify those particles at later times which belonged to one
of the two most massive progenitors before the merger. To that purpose we
apply a hierarchical  friends-of-friends approach (FoF, see
\citealt{Klypin99}) at $z=1.2$, which is $\sim2\Gyr$ before the actual
merger takes place. First the three most massive haloes within the
high resolution Lagrangian volume are found by a linking length of $ll
= 0.17$ times the mean particle separation. With a sequence of
decreasing linking lengths, namely 0.10, 0.07 and 0.02 times the mean
particle separation, substructures of increasing density are
detected. Tab.~\ref{tab:memb} lists the three most massive structures
for the various linking lengths. 

We find that the three most massive groups detected for each linking
length are associated one by one with the three most massive spatially
separated 0.17 haloes. Visual inspection reveals that the second most
massive progenitor is in the final state of a merging event at $z=1.2$.  
However, at $z\approx0.85$ when the central parts of the two progenitors
begin to penetrate each other, the relaxation associated with that
precedent merging event is entirely completed. Therefore, the two
progenitors can be assumed to be fairly undisturbed objects at the
onset of the major merger. In the following we will refer to the two most
massive progenitors, comprising 153916 and 140898 particles, as the
first and second ranked progenitors. The difference in mass is only
$\sim9$ per cent, which justifies the perception of an equal mass
merging event. To get an idea of the constellation relative to each
other we apply a further FoF analysis this time with an enlarged
linking length of 0.20 times the mean particle separation listed in
the first line in Tab.~\ref{tab:memb}. The most massive 0.20 group
embeds the two most massive  0.17 haloes, indicating that those two
are about to merge. The second 0.20 group is much less massive and
encloses the third 0.17 group, indicating that the third 0.17 group is
well separated from the two others. In fact this group will merge with
the main cluster at $z\approx0.1$, a long time after the completion of
the major merger. To summarise, at $z=0.85$ the two most massive
progenitors are fairly relaxed systems. The second ranked progenitor
shows disturbances due to a precedent merging event at $z\gtrsim1.2$,
however these are completely  erased at $z=0.85$.  The two progenitors
have similar masses and are about to merge. The third most massive
object is clearly  separated at $z=1.2$ and merges with the main
cluster not before $z=0.1$.  

\begin{figure}
\epsfig{file=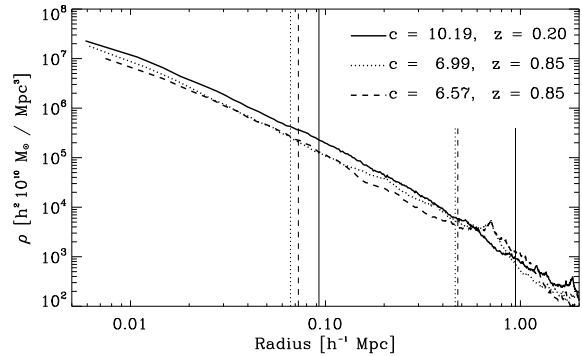,width=0.95\hsize}  
\caption{\label{fig:pro3}
Density profiles of the merged cluster at $z=0.2$ and the two most
massive progenitors at $z=0.85$. The inner vertical lines display the
characteristic radii of the NFW-fits and the outer vertical lines show
the virial radius. The humps seen in both progenitor profiles at
$\sim700\hkpc$ indicate the locations of the other progenitor at that
redshift (compare to Fig.~\ref{fig:d0.025_0217ll0.170}). 
}
\end{figure}
Fig.~\ref{fig:pro3} compares the density profiles of the first and
second ranked progenitors at $z=0.85$ with the density profile of the merged
system at $z=0.2$. In agreement with the FoF analysis, which indicated
that the second ranked progenitor formed later, the first
ranked progenitor shows higher concentration ($c=6.99$) than the
second one ($c=6.57$), where the concentration results from a NFW-fit
\citep{Navarro97} to the density profiles. The densities within the
merged system are about two times larger than the densities within the
two most massive progenitors. The hump seen in both progenitor
profiles at $r\sim700\hkpc$ reflects the location of the other
progenitor.

Simulations provide an opportunity to trace the evolution
of single particles. This is of particular interest since it enables
us to analyse the properties of the subset of particles within the merged
cluster that originate from each progenitor. In
addition the orbits of the progenitor remnants within the already
merged system can be easily traced if the substructure identification
is based only on those particles which originate from the
corresponding progenitor. Therefore, we keep record of the particles
which belonged to one of two most massive haloes at $z=1.2$. That way
the progenitor of a given particle can be identified for any given
subsequent time. A particle can either originate from the most or the
second massive progenitor at $z=1.2$ or from neither of them.

\begin{figure}
\epsfig{file=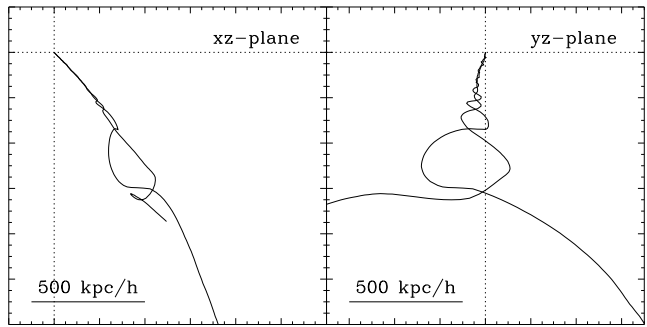,width=1.0\hsize}  
\caption{\label{fig:orbitll0.170}
The orbits of the progenitor cores projected onto the xz-plane (left
panel) and the yz-plane (right panel). The orbits are plotted between
$z\sim0.9$ and $z\sim0.3$.  
}
\end{figure}
\begin{figure}
\epsfig{file=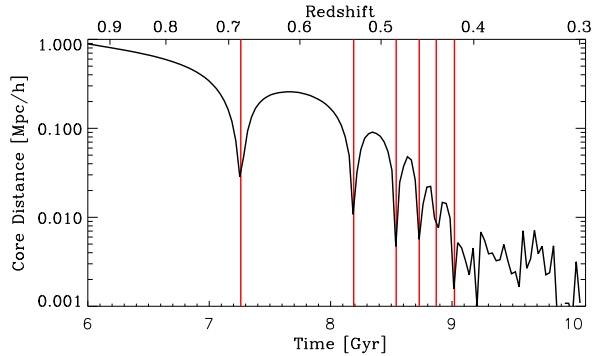,width=0.95\hsize}  
\caption{\label{fig:d0.025_0217ll0.170}
Evolution of the distance between the two progenitor cores with
time. The vertical lines indicate the 6 succeeding peri-centre
passages.}
\end{figure}

To trace the orbit of a progenitor during the merging process we
compute the top hat smoothed density field, $r_{tophat}=25\hkpc$
(proper), based only on those particles which belonged to the
corresponding 0.17 FoF group at $z=1.2$. Subsequently, we determine the
location of the peak density. Applied to all snapshots, this procedure
enables us to trace the orbit of the progenitor core during peri-centre
passages and at late evolutionary states even if there is a strong
contamination of particles from the other
progenitor. Fig.~\ref{fig:orbitll0.170} displays the projected orbits
of the two most massive progenitors between $z\sim0.9$ and $z\sim0.3$. A
complementary aspect is provided by Fig.~\ref{fig:d0.025_0217ll0.170}
which shows the evolution of the distance between the two progenitor
cores with time. During the first peri-centre passage the minimum
distance of the progenitor cores is $\sim30\hkpc$ and it gradually
declines to $\sim2\hkpc$ for the last resolved peri-centre passage. 
Similar decays of subhaloe orbits have been shown by \cite{Kravtsov04}
and \cite{BoylanKolchin06}. Our approach enables to trace six peri-centre
passages before the two progenitor cores entirely fuse to one single
structure. About $2\Gyr$ elapse between the first peri-centre passage
and the final coalescence. 
\section{Results}
\label{sec:res}
The present analysis of an N-body merging event takes advantage of the
possibility to mark particles according to their
association with one of the two most massive progenitors at $z=1.2$. 
Consequently the properties of all the particles within the final halo
volume can be compared to the properties of only those particles that
originally were associated with either the first or the second ranked
progenitor. In the foregoing section we have verified that the 
masses of the two progenitors at the onset of the merging process
($z\approx0.85$) are in fact very similar but the concentrations
differ. The first ranked progenitor is somewhat more
concentrated. This property is important for the interpretation of the
results presented here. The following paragraphs focus on the
evolution of mean values of the density, velocity dispersion and the
entropy within a variety of proper halo-centric radii ($r=25$, 50,
100, 250, 500 and $1000\hkpc$). Finally we discuss the oscillations of
the central potential during the merging process. 
\subsection{Density}
\begin{figure*}
\epsfig{file=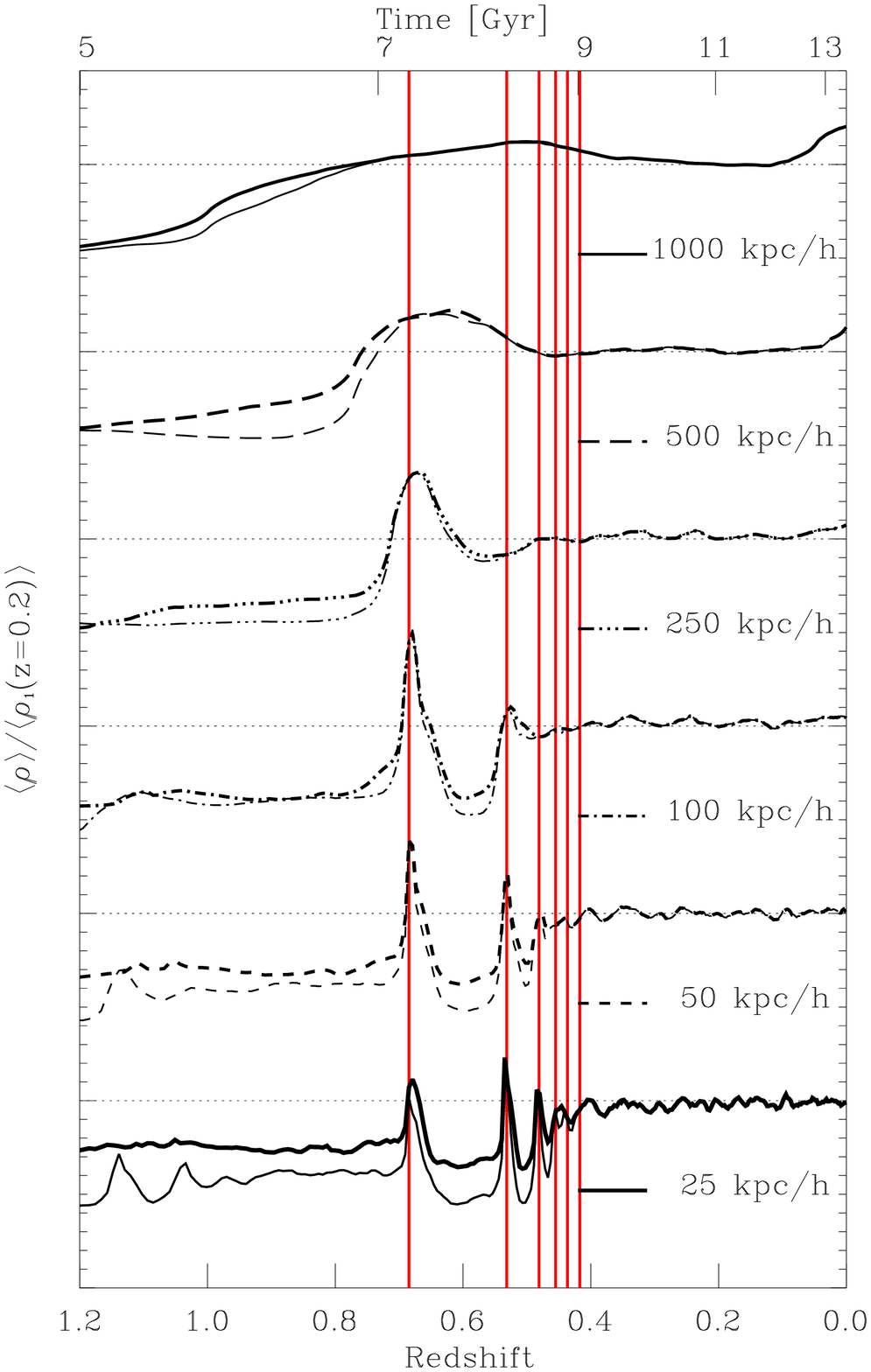,width=0.45\hsize}
\hspace{0.05\hsize}  
\epsfig{file=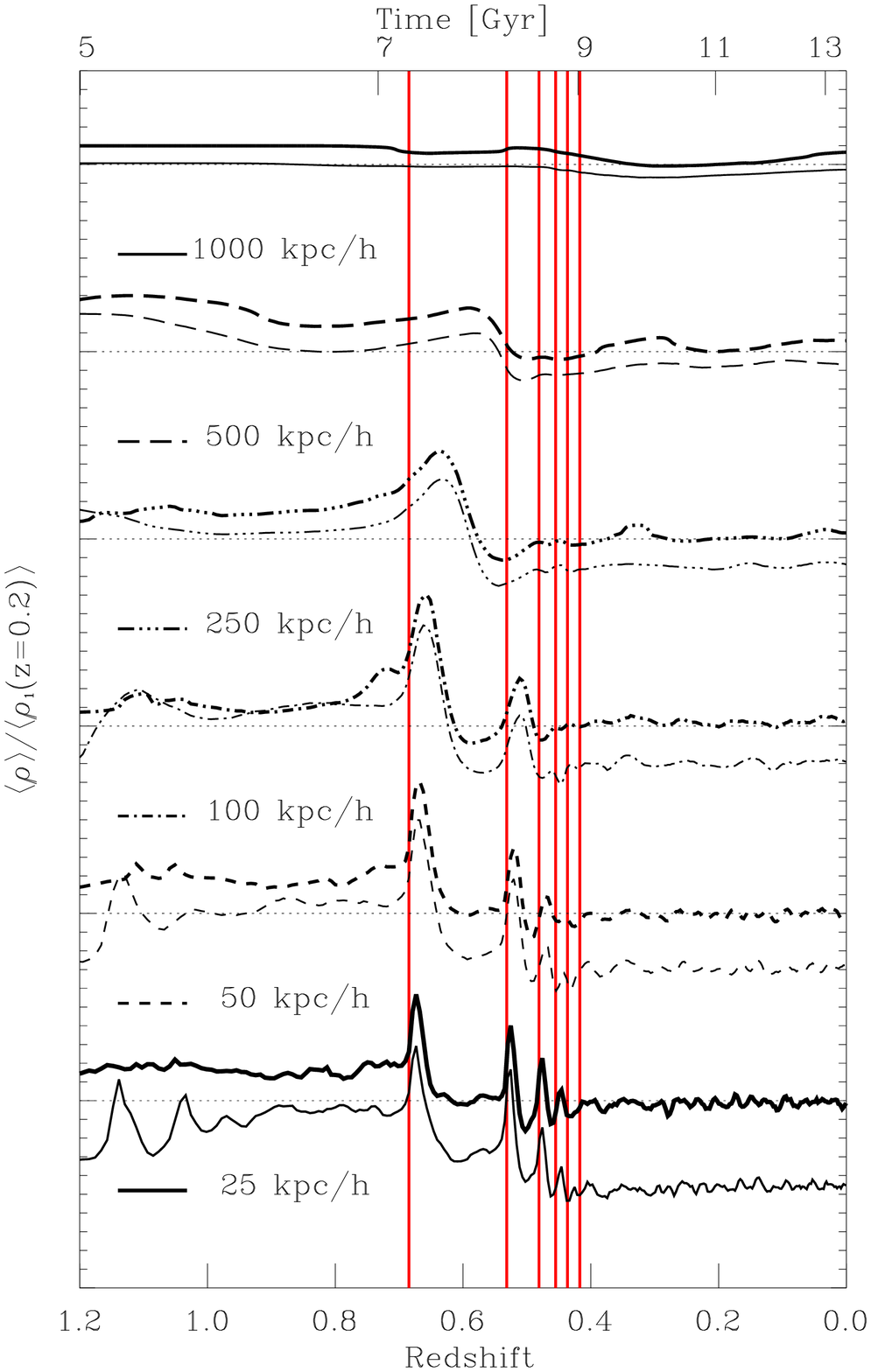,width=0.45\hsize}  
\caption{\label{fig:c0.025_0217ll0.170}
{\it Left}: Evolution of the mean densities for the two progenitors
within the listed radii scaled by the mean densities of the first
ranked progenitor $\langle\rho_1\rangle$ at $z=0.2$. The horizontal
dotted lines mark a value of 1 for the pair of graphs next to it and
a value of 0 for the pair of graphs above. The heavy lines trace the
evolution of the first ranked, more concentrated progenitor. The thin
lines display the evolution of the second ranked, less concentrated
progenitor.The vertical lines indicate the 6 succeeding peri-centre 
passages as introduced in Fig.~\ref{fig:d0.025_0217ll0.170}. 
{\it Right}: Evolution of the mean densities for the two progenitors
within the listed radii based only on the particles which were
associated with the related progenitor at $z=1.2$. The densities are
scaled by the mean densities of the first ranked progenitor
$\langle\rho_1\rangle$ at $z=0.2$. Heavy lines indicate the evolution
of the first ranked, more concentrated progenitor and thin display the
behaviour of the second ranked, less concentrated progenitor. The
vertical lines indicate the 6 succeeding peri-centre passages as
introduced in Fig.~\ref{fig:d0.025_0217ll0.170}.
}
\end{figure*}
The evolution of local densities during a merging event of
collisionless systems depends on two competing processes: (1) the
volume overlap of the two systems causes an increase of the local
densities; (2) the deepening of the gravitational potential well
that induces a contraction accompanied by accretion of additional
material from the surrounding. To disentangle these two mechanisms
Fig.~\ref{fig:c0.025_0217ll0.170} shows the evolution of the mean  
densities in two different ways. The left hand side of
Fig.~\ref{fig:c0.025_0217ll0.170} displays the evolution of the
'total' densities, i.e. densities based on all particles within the
corresponding halo-centric spheres independent of the particles'
origin (first or second ranked progenitor or none of them). The
right hand side of Fig.~\ref{fig:c0.025_0217ll0.170} 
exhibits the densities based only on those particles which originally (at
$z=1.2$) were associated with each progenitor. In both
figures the heavy lines are used for the density evolution associated
with the first ranked progenitor and the thin lines depict the
evolution of the second ranked progenitor. All densities are scaled by
the mean densities of the first ranked progenitor
$\langle\rho_1\rangle$ at $z=0.2$. (In fact, if the densities are
computed based on all particles within a sphere
$\langle\rho_1\rangle=\langle\rho_2\rangle$ since two progenitor cores
coincide at $z=0.2$). Each horizontal dotted line indicates a value of
one for the pair of graphs next to it and a value of zero for the pair
of graphs above. 

For all proper inclusion radii the evolution of the mean densities in
the left panel of Fig.~\ref{fig:c0.025_0217ll0.170} passes a maximum
value around $z=0.7$. For all radii $lesssim500\hkpc$ the mean
densities after the completion of the merging are somewhat lower than
the sum of two the corresponding progenitor densities. For example the
tracks of the mean densities within $100\hkpc$ raise from $\sim0.6$
for both progenitors to $\sim1.5$ at the first peri-centre
passage. After passing a subsequent minimum and a less pronounced
second maximum the density of the merged system settles at a value of
1, whereas  the sum of the two initial densities results in a value of
1.2 . In contrast to the final density, the density at first peri-centre
passage ($\sim1.5$) is larger than the sum of the two initial
densities. This additional increase cannot be explained solely by the 
volume overlap of the two progenitors, it must be at least partly
caused by the deepened potential well during the peri-centre passage
which attracts an additional amount of matter towards the centres of
the progenitors. 

The right hand side of Fig.~\ref{fig:c0.025_0217ll0.170} measures the
density enhancement as a response of the deepened potential well
during peri-centre passages. Here the computation of the densities is
based only on the particles which were associated with each
corresponding progenitor at $z=1.2$. This approach rules out
increasing densities as a result of overlap with the other
progenitor. The only way that the mean densities within the fixed
proper radii can increase is by contraction, causing additional
material to stream inwards across the boundary of that particular
sphere. The densities are normalised by the mean densities associated
with the first ranked progenitor $\langle\rho_1\rangle$ at $z=0.2$. In
all cases the mean densities  are lowered after the merging event and
the densities of the second ranked and less concentrated 
progenitor decrease by somewhat higher fractions. Some of the kinetic
bulk energy of the progenitors is used to accelerate individual
particles which subsequently leave the original volume. The modest
decrease for the $1000\hkpc$ is explained by the fact that before
merging the two most massive progenitors have virial radii
$\sim500\hkpc$ (see Fig.~\ref{fig:pro3}), similarly the extend of the
0.17 FoF groups at $z=1.2$ is substantially smaller than the $1000\hkpc$
spheres. Therefore, if particles are pushed towards larger radii
during the merging event they may still reside within the $1000\hkpc$ 
sphere. However even for the $1000\hkpc$ spheres a 10 percent loss can
be observed.  

It is of interest to see how much additional matter (i.e. particles)
not associated with either of the two most massive progenitors
at $z=1.2$, can be found within the various spheres after the merging
event. In the following table we display the sum of the particles
descending from both most massive progenitors divided by the total
number of particles within the listed radii. 
\begin{center}
\begin{tabular}{ccccccc}
$\hkpc$ &25   &50   &100  &250  &500  &1000 \\\hline
$(N_1+N_2)/N$ & 1.00 & 0.99 & 0.95 & 0.84 & 0.68 & 0.52
\end{tabular}
\end{center}
The table above indicates that all particles within the central
$25\hkpc$ and 95 per cent of the particles within $100\hkpc$ descend
from one of the two most massive 0.17 FoF groups at $z=1.2$. Within the
$1000\hkpc$ sphere $\sim50$ per cent of the material was not
associated with either of the two progenitors. As discussed above this is
mainly caused by the fact that the $1000\hkpc$ sphere is substantially
larger than the FoF groups, therefore already at $z=1.2$ a
substantial fraction of particles is included which is not associated
with either of the two 0.17 FoF groups.

The vertical lines in Fig.~\ref{fig:c0.025_0217ll0.170} indicate the
moments of peri-centre passage derived from the
evolution of the core distances in Fig.~\ref{fig:d0.025_0217ll0.170}. 
The right hand side of Fig.~\ref{fig:c0.025_0217ll0.170} reveals a
slightly delayed peak of the densities based only on the original
progenitor particles compared to the densities based on all particles
within the given radii. The density enhancement due to the overlap of
the two progenitors is followed by a density enhancement due to a
contraction of the original particles. For example the first peaks for
the $50\hkpc$ sphere on the right hand side of
Fig.~\ref{fig:c0.025_0217ll0.170} are delayed by $\sim100\Myr$
compared to the corresponding peaks on the left hand side. For larger
inclusion radii this delay time increases. As clarified in
Fig.~\ref{fig:d0.025_0217ll0.170} the $25\hkpc$ spheres of 
the two progenitors overlap only partially during the first
peri-centre passage with a minimum core distance of
$\sim30\hkpc$. Thus most of the density increase within the 
innermost sphere must be caused by contraction involving mostly 
original particles, which should erase differences between the
two panels. Despite of the slightly higher density peak for 
the second ranked progenitor (left panel) and a marginal
delay for both peaks of the individual progenitors (right panel), the
amplitude and shape of the first peaks of the $25\hkpc$ spheres are
very similar in both panels.   

The evolution of the densities has revealed that the
peak densities during peri-centre passages within fixed proper radii are
caused by the volume overlap and additionally by contraction as a
response of the deepened potential well during that phase. The final
masses included within the fixed radii are not the sum of the masses
within the corresponding radii of the progenitors. The transformation
of kinetic bulk energy into random velocities pushes individual
particles out of the original volume. Despite having very similar masses,
the lower concentrated progenitor loses a higher fraction of its
original central particles compared to the higher concentrated
progenitor.
\subsection{Velocity dispersion}
\begin{figure*}
\epsfig{file=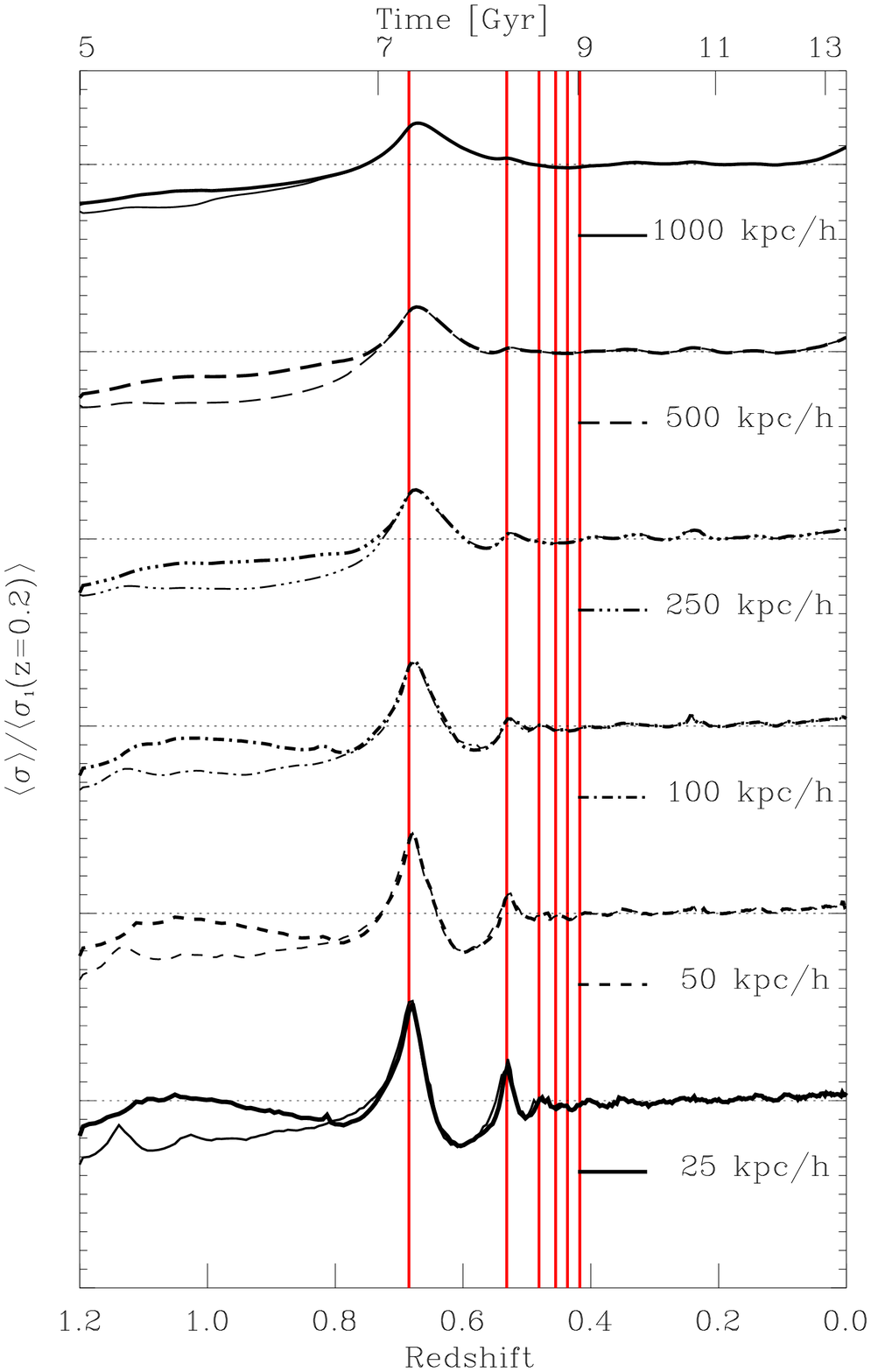,width=0.45\hsize}  
\hspace{0.05\hsize}  
\epsfig{file=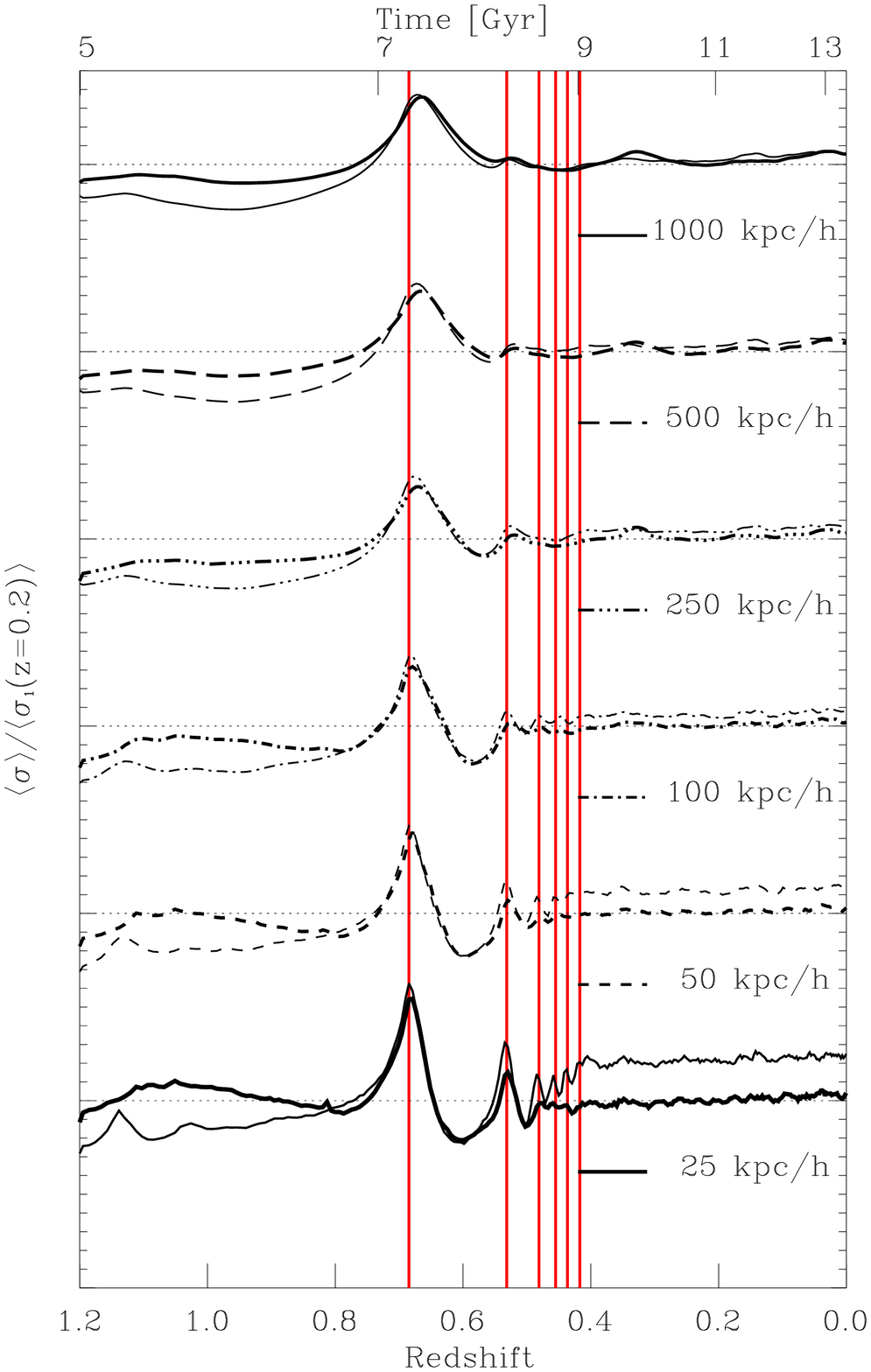,width=0.45\hsize}  
\caption{\label{fig:vc0.025_0217ll0.170}
{\it Left}: Evolution of the mean velocity dispersions for the two
progenitors within the listed radii scaled by the values associated
with the first ranked progenitor $\langle\sigma_1\rangle$ at $z=0.2$.  
The interpretation of the lines and line-styles is the same as in
Fig.~\ref{fig:c0.025_0217ll0.170}. {\it Right}: Evolution of the mean
velocity dispersions for the two progenitors within the listed radii
based only on the particles which were associated with the related
progenitor at $z=1.2$. The dispersions are scaled by the mean values
associated with first ranked progenitor $\langle\sigma_1\rangle$ at
$z=0.2$. The interpretation of the lines and line-styles
is the same as in Fig.~\ref{fig:c0.025_0217ll0.170}.
}
\end{figure*}
Using an approach similar to that of the densities also the mean
velocity dispersions are investigated in a twofold
manner. Fig~\ref{fig:vc0.025_0217ll0.170} shows the evolution of the
mean velocity dispersion for all particles within the corresponding
spheres and for only those particles which were associated with each
individual progenitor at $z=1.2$. The heavy lines display
the evolution of the first ranked progenitor whereas the
thin lines display the fate of the second ranked progenitor. 
Before computing the velocity dispersions the bulk 
velocity has to be subtracted from the individual particle
velocities. For the the bulk velocity of the merging subhaloes we
adopt the mean velocity within the central $25\hkpc$ sphere,
considering  only those particles associated with each 
progenitor at $z=1.2$. This approach avoids a high background
contamination, in particular during peri-centre 
passages. In addition the small volume sphere ($r=25\hkpc$) allows us
to focus on the very central processes. As a test we also computed
the evolution of the velocity dispersions based on bulk velocities
within $250\hkpc$ spheres resulting in negligible
deviations. The red vertical lines depict the moments of the
peri-centre passages obtained from Fig.~\ref{fig:d0.025_0217ll0.170}.

We find the velocity dispersions within fixed proper radii do not
change by more than 20 per cent between the values before and after
the merging event. This is valid for the total velocity dispersion
displayed in the left panel of Fig.~\ref{fig:vc0.025_0217ll0.170} as
well as for the velocity dispersions based only on individual progenitor
particles on the right. During the first peri-centre 
passage the velocity dispersions exhibit a momentary increase of
$\lesssim60$ per cent. At the subsequent apo centre passage the
velocity dispersions recede to the pre-merging values. The 
dispersions within the $25\hkpc$ sphere drop even slightly below the
initial values. At the following peri-centre passage a second peak
with substantial lower amplitude is reached. The following
oscillations for the total velocity dispersions indicate some
relaxation process but a synchronisation with the peri-centre
passages is ambiguous. A slightly different picture arises for
dispersions based only on the original progenitor particles (right
panel of Fig.~\ref{fig:vc0.025_0217ll0.170}). At least
for the central dispersions ($r\leq100\hkpc$) the particles descended
from the second ranked progenitor show a synchronisation with the
peri-centre passages.

The double peak structure in both panels of
Fig.~\ref{fig:vc0.025_0217ll0.170} in particular for the spheres with
intermediate radii resembles the behaviour of the total kinetic energy 
found by \cite{Pearce93} who analysed isolated N-body mergers. 
However, one has to keep in mind that our approach probes the
evolution of the mean velocity dispersion along the paths of the
progenitors which is not exactly the same as the total
kinetic energy of the whole system. Nevertheless, the fact that we find
a similar double peak structure reveals that the internal velocity
dispersions have a strong impact on the total kinetic energy. The 
structural similarity of the initial peaks in the left and the
right panel of Fig.~\ref{fig:vc0.025_0217ll0.170} indicates that the
increase in velocity dispersion is mainly caused by an internal  
acceleration due to the deepening potential well during
peri-centre passages. This statement is supported particularly by the
evolution of the velocity dispersions within the $25\hkpc$ spheres. 
As shown in Fig.~\ref{fig:d0.025_0217ll0.170} the distance of the
cores is $\sim30\hkpc$ resulting in only a partial overlap of the
spheres associated with the two progenitors. Therefore, the enhancement
of the total velocity dispersion is based mainly on the original
particles of the associated progenitor and should resemble the behaviour
in the right hand panel, as it does. However, a slight difference
between the evolution of the total velocity dispersions and those based
only on original progenitor particles can be seen for almost all
initial peaks. The peak associated with the second ranked progenitor
(thin lines in the right panel) is slightly higher than the peak for the
particles descending from the first ranked progenitor. For the central
spheres this trend is more pronounced during the subsequent peri-centre
passages, resulting in substantially larger velocity dispersions for
the particles descending from second ranked progenitor after total
coalescence (thin lines on the right panel of  
Fig.~\ref{fig:vc0.025_0217ll0.170}). 

Combining the results derived for the evolution of the densities
with the findings of this section results in the following
picture. After completion of the merging process the densities based on
the particles descending from the second ranked progenitor are lowered
relative to the densities related to the first ranked progenitor. The
velocity dispersions behave in the opposite way, they are higher for
the particles descending from the second ranked progenitor. 
The progenitors have nearly equal masses but different
concentrations, which motivates the conclusion that an equal mass
merger transfers a larger fraction of the released gravitational energy 
to the system of lower concentration and/or the energy
is distributed amongst fewer particles in the low concentrated
system. This behaviour is probably related to the findings by
\cite{Dehnen05} who showed based on the phase space analysis of
merging systems that the steepest cusps survive. However, it is not
clear whether there exists a correlation between the central slope and
concentration (see e.g.~\citealt{Diemand04}).      
\subsection{Velocity anisotropy parameter}
\begin{figure*}
\epsfig{file=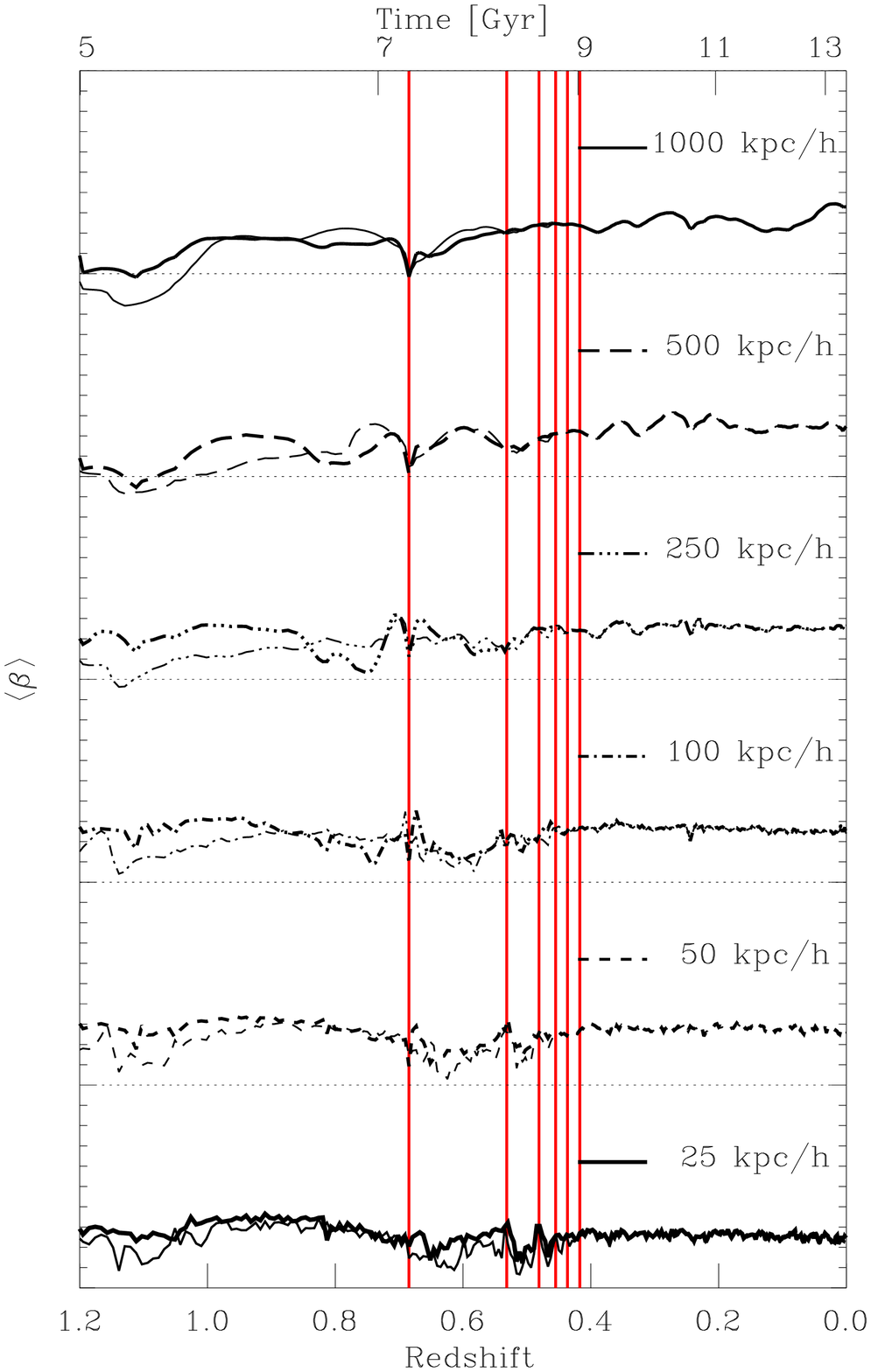,width=0.45\hsize}  
\hspace{0.05\hsize}  
\epsfig{file=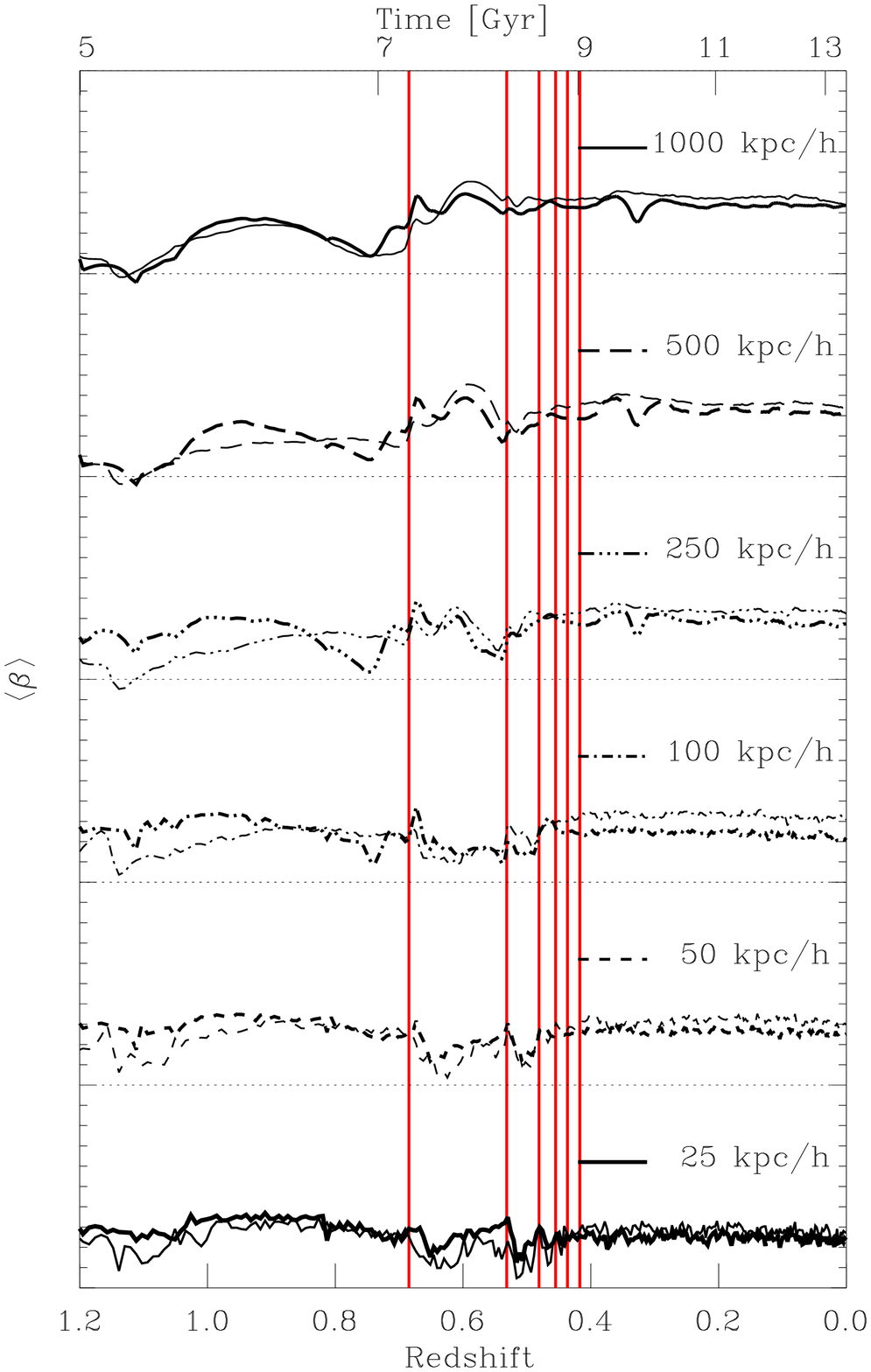,width=0.45\hsize}  
\caption{\label{fig:bec0.025_0217ll0.170}
{\it Left}: Evolution of the mean velocity anisotropy parameter
$\beta$ for the two progenitors within the listed radii. The
velocities of all the particles within the corresponding radii are
taken into account. Heavy lines display the evolution of the first
ranked progenitor and thin lines represent the second ranked
progenitor. {\it Right}: Evolution of the mean velocity anisotropy
parameter $\beta$ for the two progenitors within the listed
radii. Only the velocities of the particles associated with the
related progenitor are considered. Heavy lines display the evolution
of the first ranked progenitor and thin lines represent the second
ranked progenitor.   
}
\end{figure*}
If during the relaxation process the particles of the lower
concentrated system are pushed further away from the central regions
this may have an impact on the velocity anisotropy of the particles
descending from this second ranked
progenitor. The left panel of Fig.~\ref{fig:bec0.025_0217ll0.170}
displays the evolution of the velocity anisotropy parameter
$\beta=1-0.5\ \sigma_t/\sigma_r$, where $\sigma_t$ indicates the
tangential velocity dispersion and $\sigma_r$ is the radial velocity
dispersion, for all particles within the corresponding spheres. At
$z=0.7$, the time of the first peri-centre passage of the 
progenitor cores, the velocities within $500$ and $1000\hkpc$ appear
to be momentary isotropic which is caused by the large
tangential contribution of the passing cores and does not reflect the
overall behaviour of the velocity fields. For smaller inclusion radii
the anisotropy drops during and a short time after the second peri-centre
passages. After $z=0.4$ a stable configuration is achieved with a
slightly radial anisotropy and a tendency to more isotopic velocities
towards the centre. 

The velocity anisotropy based only on the particles descending from
one of the progenitors at $z=1.2$ (right panel of
Fig.~\ref{fig:bec0.025_0217ll0.170}) support the above discussion of 
the velocity dispersions. After the final coalescence of the
progenitor cores the particles related to the second ranked progenitor
show higher radial anisotropies than particles associated with the first
ranked progenitor. The analysis of the densities and the velocity
dispersions have already revealed that the particles descended from
the less concentrated progenitor are more strongly dispersed. To make
that happen these particles have to be provided with some additional
radial velocity component which is supplied by the
gravitational energy released by of the merging systems. The result of
this process is an relative increase of the radial anisotropy causing
higher values of the $\beta$ parameter. 
\subsection{Entropy}
\begin{figure*}
\epsfig{file=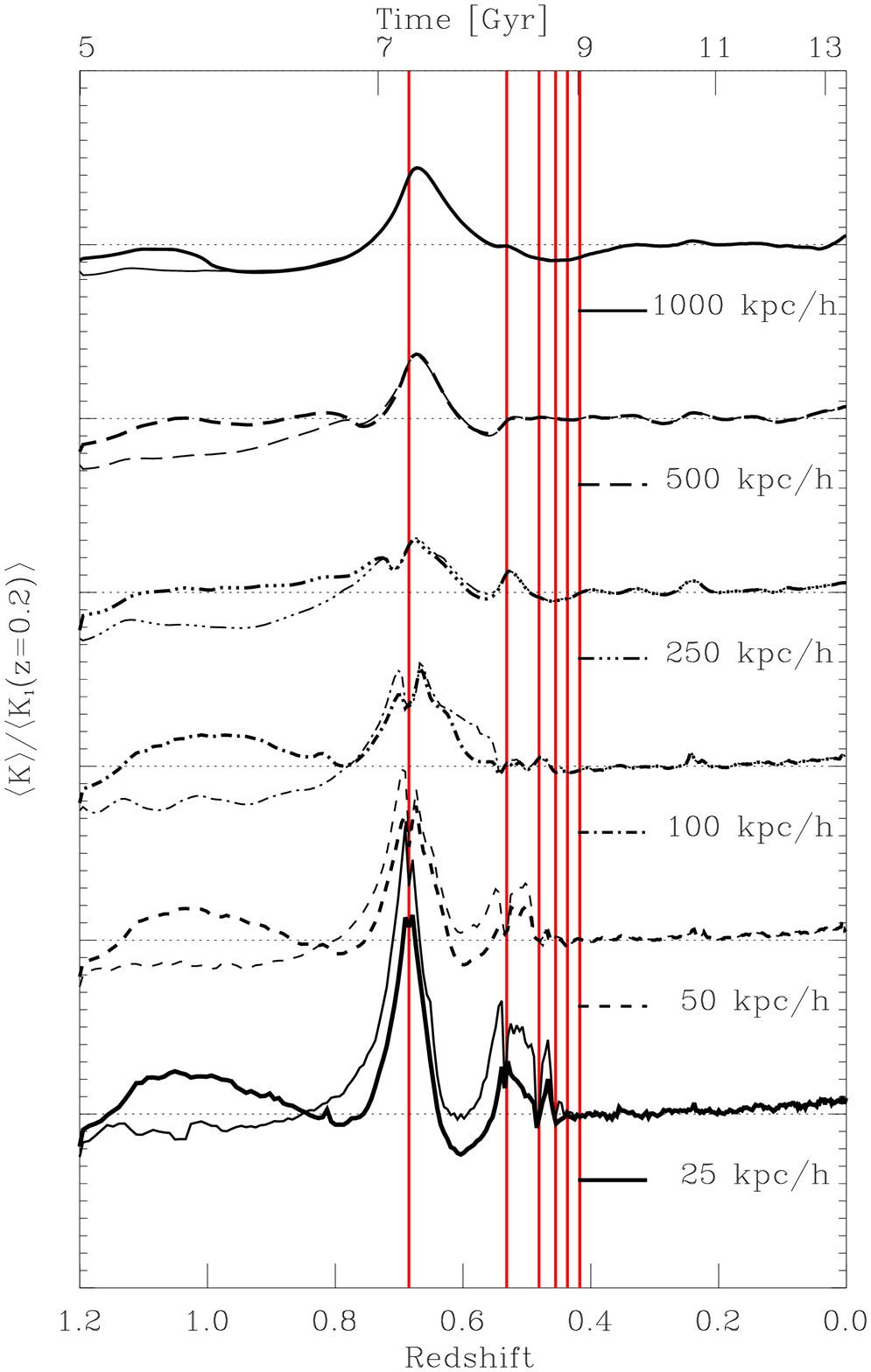,width=0.45\hsize}  
\hspace{0.05\hsize}
\epsfig{file=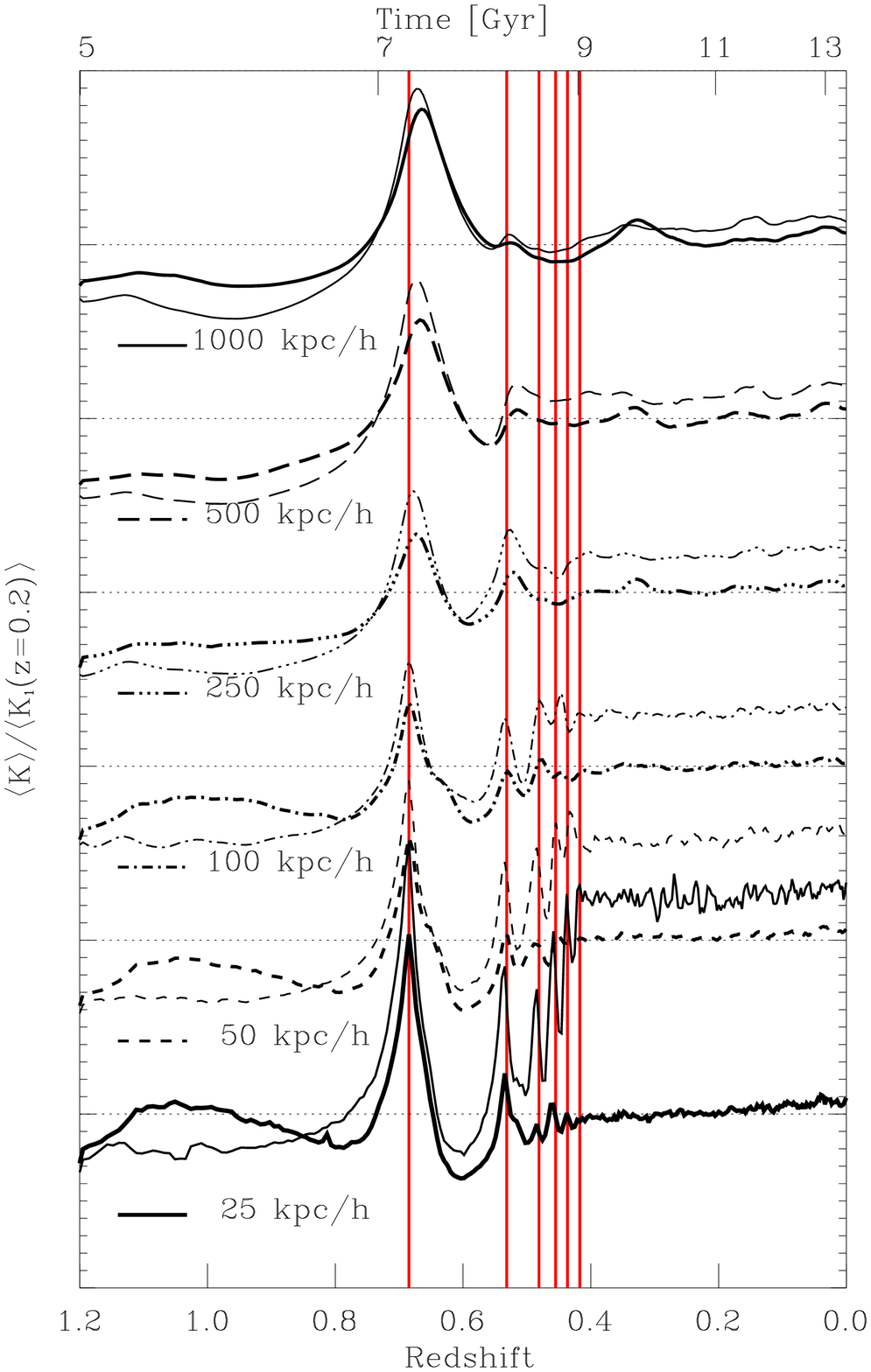,width=0.45\hsize}  
\caption{\label{fig:kc0.025_0217ll0.170}
{\it Left}: Evolution of the mean entropies ($K=\sigma^2/\rho^{2/3}$)
for the two progenitors within the listed radii scaled by the values
associated with the first ranked progenitor $\langle K_1\rangle$ at
$z=0.2$. The interpretation of the lines and line-styles is the 
same as in Fig.~\ref{fig:c0.025_0217ll0.170}. {\it Right}: Evolution
of the mean entropy ($K=\sigma^2/\rho^{2/3}$) for the two progenitors
within the listed radii based only on the particles which were
associated with the related progenitor at $z=1.2$. The mean entropies
are scaled by the values associated with the first ranked progenitor
$\langle K_1\rangle$ at $z=0.2$. The interpretation
of the lines and line-styles is the same as in
Fig.~\ref{fig:c0.025_0217ll0.170}.   
}
\end{figure*}
\begin{figure*}
\epsfig{file=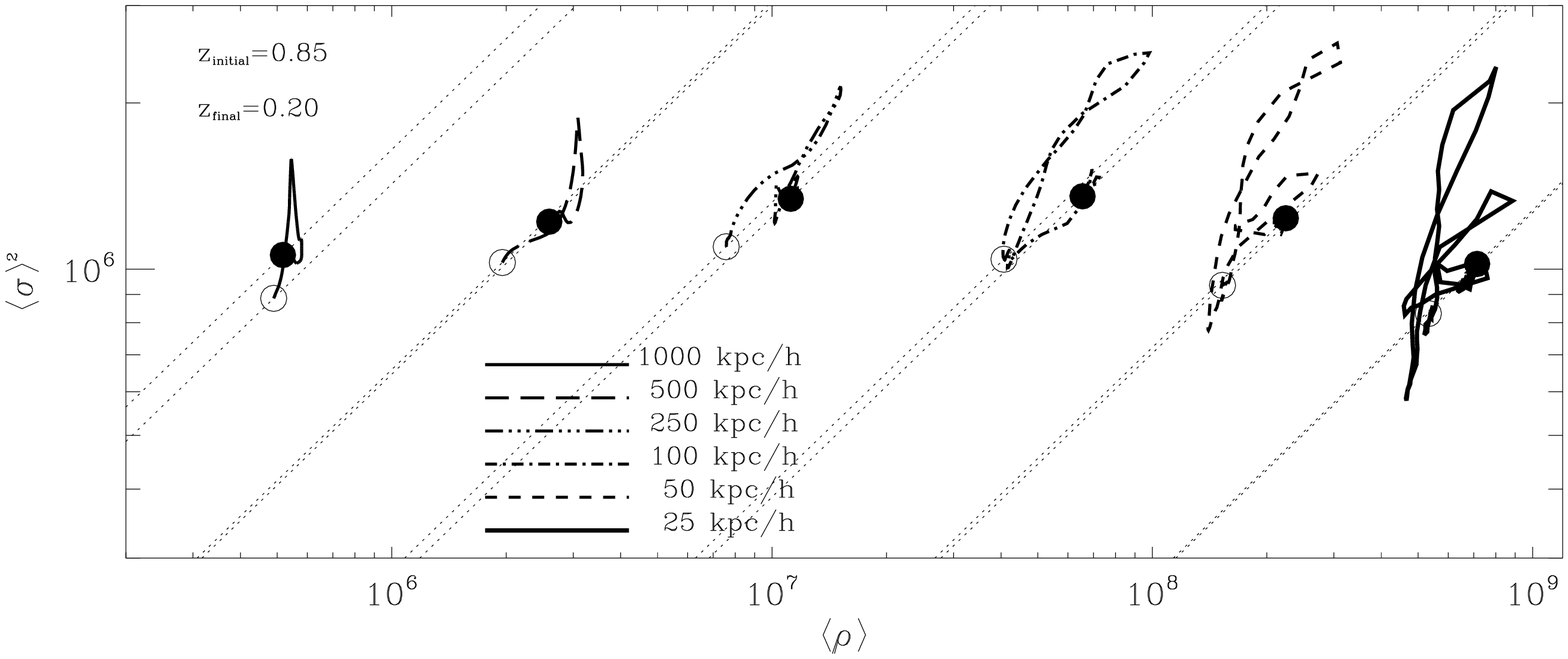,width=0.95\hsize}  
\caption{\label{fig:sdc0.025_0217ll0.170}
Evolution of the first ranked progenitor for the listed inclusion
radii in the density-velocity dispersion plane. The slanted lines show
loci of constant entropy $K=\sigma^2/\rho^{2/3}$ of equivalently
phase-space density $\rho/\sigma^3$. The tracks begin at a redshift of
$z=0.85$ (approximately the beginning of the merging event) marked by
open circles and end at a redshift of $z=0.2$ after the progenitor
cores have merged indicated by filled circles. Units are arbitrary and
relative values of the entropy increase toward the upper left. Compare
with Fig. 8 of Navarro et al. (1995). 
}
\end{figure*}
We define the entropy of the DM similar to the entropy of an ideal gas
(see e.g. \citealt{White87, Navarro95, Eke98}). In analogy to the
Sackur-Tetrode equation we write for the specific
entropy of the dark matter $S_{\rm DM}=\ln(\sigma{_{\rm DM}^{3/2}} /
\rho_{\rm DM})$, where $\sigma_{\rm DM}$ is the DM velocity dispersion and an
additive constant is neglected. Following the convention used in the
studies of galaxy clusters we define $K_{\rm DM}=\sigma{^2_{\rm
DM}}\rho{^{-2/3}_{\rm DM}}$ (see also \citealt{Faltenbacher06b}). The 
DM specific entropy  is closely related to the phase space density
$Q_{\rm DM}=\rho_{\rm DM} \sigma_{\rm DM}^{-3}$ \citep{Taylor01,
Ascasibar04}.    

Fig.~\ref{fig:kc0.025_0217ll0.170} displays the mean entropies within
the listed radii along the tracks of the two most massive
progenitors. During the merging event the entropy within all radii
rises temporarily. But it is difficult to estimate the change from 
initial to final entropy since the initial values are ill
determined, e.g. the central entropy of the first ranked progenitor
(thick lines) shows a maximum value at $z\sim1$ and drops by $\sim20$
per cent before dramatic growth takes place during the first
peri-centre passage. In Fig.~\ref{fig:sdc0.025_0217ll0.170} we show
evolution within the density-velocity dispersion plane between
$z=0.85$ and $z=0.2$. The open circles indicate the starting point of
the tracks and the filled circles give the location at $z=0.2$.  
Attached to both of these points is a dotted line indicating the loci
of constant entropy or equivalently phase-space density. The distance
between a pair of dotted lines gives the ratios between the final
and initial entropies $K_f/K_i$, which range from $0.95$ to $1.15$ for
the $100\hkpc$ and the $1000\hkpc$, respectively. This changes are
rather small, in particular if one considers the occasionally sever
deviations from isentropic behaviour during the active merging phase. 
Within the $500\hkpc$ sphere initial and final entropies are equal
(heavy long dashed line in Fig.~\ref{fig:kc0.025_0217ll0.170}). This
radius may be of particular interest, since it is the largest radius
included by the virial radius for all times considered here.   

For the 100 and $250\hkpc$ spheres the evolution of the entropies show a
clear double-peaked structure at the first peri-centre passage. This
feature is most prominent for the $100\hkpc$ sphere. The analysis of the 
the densities (Fig.~\ref{fig:c0.025_0217ll0.170}) revealed that
within this sphere the density during the first peri-centre passage
increases strongest compared to the behaviour of other spheres. This
particular steep density enhancement can counterbalance the increase
of the entropy caused by the peaking velocity dispersion. 

We have shown that particles descended from the second
ranked progenitor build a subset of particles within the final cluster
with lower central densities and higher velocity dispersions compared
to particles descended from the first ranked, more concentrated
progenitor. The combination of these two processes causes a substantial
rise in the central entropies for the particles descended from the
second ranked progenitor as displayed in
Fig.~\ref{fig:kc0.025_0217ll0.170}. This is a further indication that
the central parts of the less concentrated progenitor get heavily
disrupted whereas the more concentrated progenitor is less affected by
the merging event. 

\subsection{Central Potential}
\begin{figure}
\epsfig{file=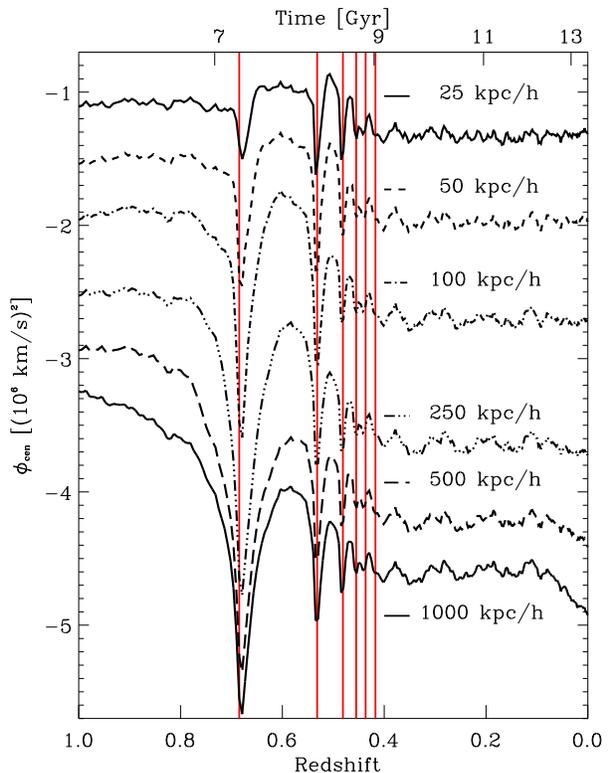,width=0.95\hsize}  
\caption{\label{fig:pc0.025_0217_0001}
Impact of the merger on the central gravitational potential 
computed within the listed radii along the track of the most massive
progenitor. 
}
\end{figure}
\begin{figure}
\epsfig{file=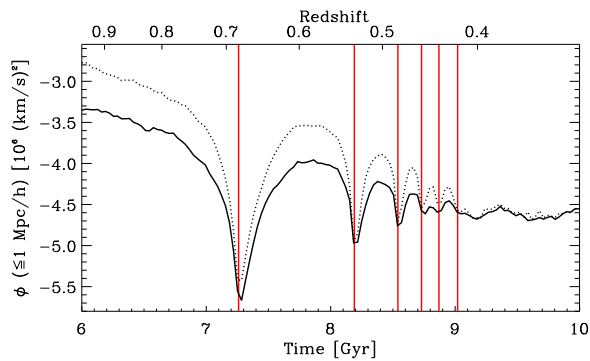,width=0.95\hsize}  
\caption{\label{fig:pc0.025_0217_0001ll0.170}
Evolution of the central gravitational potential computed within 
the $1\hMpc$ centred along the tracks of the two most massive
progenitors. The solid line displays the evolution of the first
ranked, more concentrated progenitor and the dotted line shows the
behaviour of the second ranked, less concentrated progenitor.
}
\end{figure}
We compute the contribution of the various spheres to the central
potential by direct summation.  
\begin{equation}
\phi(\leq r) = - \G \sum_{N(\leq r)} \frac{m_p}{r_i - r_{cen}}
\end{equation}
Here, $r_{cen}$ is the centre of the progenitor, $m_p$ and $r_i$ are
the mass and the position of the $i$th particle within the inclusion
radius $r$ and $\G$ is the gravitational constant. The evolution for 
the central potential of the most massive progenitor is displayed in
Fig.~\ref{fig:pc0.025_0217_0001}. In order to see how the evolution of
the potential is affected by occasional very central passages of
single particles we have computed the potential of the particle
which closest to the centre for the given moment.  
We find that the contribution of the most central particle to the
potential is on average $-0.003\times 10^6 (\kms)^2$ with maximal
amplitudes of $\sim-0.01\times 10^6 (\kms)^2$. Therefore, occasional
central passages of single particles do not seriously affect the
overall evolution of the gravitational potential. The oscillatory
behaviour seen in the evolution of the density and the velocity
dispersion is also apparent in the central potential. Since the
potential based on the $1000\hkpc$ results in the most pronounced
variations we compare in Fig.~\ref{fig:pc0.025_0217_0001ll0.170} the
central potentials along the tracks of the two progenitors based on
the $1000\hkpc$ sphere. Obviously the central potential of the
second ranked progenitor undergoes stronger variations during the 
oscillatory relaxation phase. The six peri-centre passages based on
the core distances (Fig.~\ref{fig:d0.025_0217ll0.170}) are clearly
visible in the evolution of the central potential. The time interval
of the later oscillations is $\sim10^8\yr$.
\section{Summary and Discussion}
\label{sec:sum}
We have analysed the merging of a cluster-sized dark matter haloe
focusing on the evolution of the density and the velocity dispersion
within fixed proper radii. This approach is complementary to the more
common analysis based on the scaling relations for dark matter haloes.
Whereas the latter emphasises the self similar nature of
gravitationally collapsed systems, the former provides insight into
the processes at fixed scales and may help to understand the cause of
the self similarity. Our main results are: 
(1) 
About $3\Gyr$s elapse from the first encounter of the progenitor
outskirts to the final coalescence of their central cores. Within  the
$\sim2\Gyr$s between the first peri-centre passage and the final core
merging six peri-centre passages take place. During that
process the periods decrease by a factor of ten from $\sim10^9\yr$s   
to $\sim10^8\yr$s. 
(2) 
The central densities of the merging progenitors peak at each
peri-centre passage and are most pronounced at the first one. The density
enhancement is caused by the overlapping volumes of the subhaloes but
also by the contraction due to the deepening of the potential well
during a peri-centre passage.       
(3)  
After the merger is completed ($z\lesssim0.4$) the densities and the
velocity dispersions within the fixed proper radii keep roughly 
constant. The increase for larger radii at $z\lesssim0.1$ is
associated with an ongoing intermediate merger as discussed in 
\S~\ref{sec:simda}. In agreement with \cite{RomanoDiaz06} we find that
the properties of the dark matter haloe change rapidly during merging
events but remain constant during phases of passive evolution.
(4) 
The ranking of the progenitor concentrations determines the final
distribution of the particles. Less concentrated haloes get more
dispersed and show higher radially anisotropic velocity dispersions. 

Finally, we briefly discuss the impacts of the oscillatory
relaxation process of the collisionless matter on the baryonic component.   
It is of interest to consider the possibility that enhanced AGN
activity in the cluster-centred galaxies may be stimulated by the
large oscillations of the haloe potentials during a major cluster
merger.  We note that the period of relaxation oscillations for our
merging cluster, $\sim 10^8$ yrs, is similar to the duty cycle
observed in AGNs \citep{Voit05}, although only a small
fraction of AGNs reside in massive clusters.  Nevertheless, to explore
this idea, we performed a simple gasdynamical calculation to determine
how much additional radiative cooling occurs when the merging
potentials are deepened, compressing the hot cluster gas.  For this
purpose we set up a traditional quasi-steady state cluster cooling
flow in a potential of mass $3 \times 10^{14}$ with typical
concentration $c = 6.74$.  The gas temperature at large radii, $3.5$
keV, was chosen according to the observed cluster $M - T$ relation
\citep{Arnaud05} and the initial (hydrostatic) gas density was
adjusted until the baryon mass matched the cosmic value at the virial
radius.  The relaxed steady state cooling flow that results closely
resembles the observations of \cite{Vikhlinin06} and has a central
cooling rate of ${\dot M} = 43$ $M_{\odot}$ yr$^{-1}$ (assuming
abundance 0.4 solar).  The potential was then set into sinusoidal
oscillations with period $0.6$ Gyr and an amplitude similar to that of
Figure 10 by sinusoidally varying the concentration between $c$ and
$0.43c$. During these oscillations, the cooling rate ${\dot M}(t)$
near the centre of the flow varied with an amplitude of only 7
per cent, much less than the unperturbed steady state value which is at
least $\sim10$ times larger than allowed by X-ray observations
\citep{Peterson01}.  We conclude that the additional radiative
cooling stimulated by the oscillating merger potentials is negligible.

Perhaps a more fruitful approach along these lines is the possibility
that galaxies (and large black holes) orbiting about the cluster core
can be processed by dynamical friction into the central galaxy/black
hole during the relaxation process.  The acceleration due to dynamical
friction $(du/dt)_{df} \propto m \rho/u^2$ depends on the mass $m$ and
the velocity $u$ of the orbiting object and the density of the ambient
material. By setting the velocity equal to the mean dispersion
$u=\sigma$, we see that the time scale for dynamical friction $t_{df}
= u/(du/dt)_{df} \propto \sigma^3/\rho = K^{2/3}$. During apo-centre
passages when the dark halo entropy passes a minimum, objects orbiting
near the cluster-centred galaxy may efficiently reduce their angular
momentum, increasing the chance of a capture by the central black hole
during the subsequent contraction phase. It is conceivable that such
events, stimulated by the global merger, could result in enhanced AGN
activity with a period imprinted by the dark matter oscillations.
\section*{Acknowledgements} 
This work has been supported by NSF grant AST 00-98351 and NASA grant
NAG5-13275 for which we are very grateful. The simulations were
performed at the Leibniz Rechenzentrum Munich. 

\begin{thebibliography}{}

\bibitem[\protect\citeauthoryear{{Arnaud}, {Pointecouteau} \& {Pratt}}{{Arnaud}
  et~al.}{2005}]{Arnaud05}
{Arnaud} M.,  {Pointecouteau} E.,    {Pratt} G.~W.,  2005, A\&A, 441, 893

\bibitem[\protect\citeauthoryear{{Ascasibar}, {Yepes}, {Gottl{\"o}ber} \&
  {M{\"u}ller}}{{Ascasibar} et~al.}{2004}]{Ascasibar04}
{Ascasibar} Y.,  {Yepes} G.,  {Gottl{\"o}ber} S.,    {M{\"u}ller} V.,  2004,
  MNRAS, 352, 1109

\bibitem[\protect\citeauthoryear{{Boylan-Kolchin} \& {Ma}}{{Boylan-Kolchin} \&
  {Ma}}{2006}]{BoylanKolchin06}
{Boylan-Kolchin} M.,  {Ma} C.-P.,  2006, ArXiv Astrophysics e-prints, {\tt
  astro-ph/0608122}

\bibitem[\protect\citeauthoryear{{Clowe}, {Bradac}, {Gonzalez}, {Markevitch},
  {Randall}, {Jones} \& {Zaritsky}}{{Clowe} et~al.}{2006}]{Clowe06}
{Clowe} D.,  {Bradac} M.,  {Gonzalez} A.~H.,  {Markevitch} M.,  {Randall}
  S.~W.,  {Jones} C.,    {Zaritsky} D.,  2006, ArXiv Astrophysics e-prints {\tt
  astro-ph/0608407}

\bibitem[\protect\citeauthoryear{{Conselice}, {Gallagher} III \&
  {Wyse}}{{Conselice} et~al.}{2001}]{Conselice01}
{Conselice} C.~J.,  {Gallagher} III J.~S.,    {Wyse} R.~F.~G.,  2001, AJ, 122,
  2281

\bibitem[\protect\citeauthoryear{{Dehnen}}{{Dehnen}}{2005}]{Dehnen05}
{Dehnen} W.,  2005, MNRAS, 360, 892

\bibitem[\protect\citeauthoryear{{Diemand}, {Moore} \& {Stadel}}{{Diemand}
  et~al.}{2004}]{Diemand04}
{Diemand} J.,  {Moore} B.,    {Stadel} J.,  2004, MNRAS, 353, 624

\bibitem[\protect\citeauthoryear{{Eke}, {Navarro} \& {Frenk}}{{Eke}
  et~al.}{1998}]{Eke98}
{Eke} V.~R.,  {Navarro} J.~F.,    {Frenk} C.~S.,  1998, ApJ, 503, 569

\bibitem[\protect\citeauthoryear{{Faltenbacher}, {Hoffman}, {Gottloeber} \&
  {Yepes}}{{Faltenbacher} et~al.}{2006}]{Faltenbacher06b}
{Faltenbacher} A.,  {Hoffman} Y.,  {Gottloeber} S.,    {Yepes} G.,  2006, ArXiv
  Astrophysics e-prints {\tt astro-ph/0608304}

\bibitem[\protect\citeauthoryear{{Klypin}, {Gottl{\"o}ber}, {Kravtsov} \&
  {Khokhlov}}{{Klypin} et~al.}{1999}]{Klypin99}
{Klypin} A.,  {Gottl{\"o}ber} S.,  {Kravtsov} A.~V.,    {Khokhlov} A.~M.,
  1999, ApJ, 516, 530

\bibitem[\protect\citeauthoryear{{Klypin}, {Kravtsov}, {Bullock} \&
  {Primack}}{{Klypin} et~al.}{2001}]{Klypin01}
{Klypin} A.,  {Kravtsov} A.~V.,  {Bullock} J.~S.,    {Primack} J.~R.,  2001,
  ApJ, 554, 903

\bibitem[\protect\citeauthoryear{{Kravtsov}, {Gnedin} \& {Klypin}}{{Kravtsov}
  et~al.}{2004}]{Kravtsov04}
{Kravtsov} A.~V.,  {Gnedin} O.~Y.,    {Klypin} A.~A.,  2004, ApJ, 609, 482

\bibitem[\protect\citeauthoryear{{Kravtsov}, {Klypin} \& {Khokhlov}}{{Kravtsov}
  et~al.}{1997}]{Kravtsov97}
{Kravtsov} A.~V.,  {Klypin} A.~A.,    {Khokhlov} A.~M.,  1997, ApJS, 111, 73

\bibitem[\protect\citeauthoryear{{Nagai}, {Vikhlinin} \& {Kravtsov}}{{Nagai}
  et~al.}{2006}]{Nagai06}
{Nagai} D.,  {Vikhlinin} A.,    {Kravtsov} A.~V.,  2006, ArXiv Astrophysics
  e-prints {\tt astro-ph/0609247}

\bibitem[\protect\citeauthoryear{{Navarro}, {Frenk} \& {White}}{{Navarro}
  et~al.}{1995}]{Navarro95}
{Navarro} J.~F.,  {Frenk} C.~S.,    {White} S.~D.~M.,  1995, MNRAS, 275, 720

\bibitem[\protect\citeauthoryear{{Navarro}, {Frenk} \& {White}}{{Navarro}
  et~al.}{1997}]{Navarro97}
{Navarro} J.~F.,  {Frenk} C.~S.,    {White} S.~D.~M.,  1997, ApJ, 490, 493

\bibitem[\protect\citeauthoryear{{Pearce}, {Thomas} \& {Couchman}}{{Pearce}
  et~al.}{1993}]{Pearce93}
{Pearce} F.~R.,  {Thomas} P.~A.,    {Couchman} H.~M.~P.,  1993, MNRAS, 264,
  497

\bibitem[\protect\citeauthoryear{{Pearce}, {Thomas} \& {Couchman}}{{Pearce}
  et~al.}{1994}]{Pearce94}
{Pearce} F.~R.,  {Thomas} P.~A.,    {Couchman} H.~M.~P.,  1994, MNRAS, 268,
  953

\bibitem[\protect\citeauthoryear{{Peterson}, {Paerels}, {Kaastra}, {Arnaud},
  {Reiprich}, {Fabian}, {Mushotzky}, {Jernigan} \& {Sakelliou}}{{Peterson}
  et~al.}{2001}]{Peterson01}
{Peterson} J.~R.,  {Paerels} F.~B.~S.,  {Kaastra} J.~S.,  {Arnaud} M.,
  {Reiprich} T.~H.,  {Fabian} A.~C.,  {Mushotzky} R.~F.,  {Jernigan} J.~G.,
  {Sakelliou} I.,  2001, A\&A, 365, L104

\bibitem[\protect\citeauthoryear{{Romano-Diaz}, {Faltenbacher}, {Jones},
  {Heller}, {Hoffman} \& {Shlosman}}{{Romano-Diaz} et~al.}{2006}]{RomanoDiaz06}
{Romano-Diaz} E.,  {Faltenbacher} A.,  {Jones} D.,  {Heller} C.,  {Hoffman} Y.,
     {Shlosman} I.,  2006, ApJ, 637, L93

\bibitem[\protect\citeauthoryear{{Spergel}, {Bean}, {Dore'}, {Nolta}, {Bennett}
  \& {et al.}}{{Spergel} et~al.}{2006}]{Spergel06}
{Spergel} D.~N.,  {Bean} R.,  {Dore'} O.,  {Nolta} M.~R.,  {Bennett} C.~L.,
  {et al.} 2006, ArXiv Astrophysics e-prints {\tt astro-ph/0603449}

\bibitem[\protect\citeauthoryear{{Taylor} \& {Navarro}}{{Taylor} \&
  {Navarro}}{2001}]{Taylor01}
{Taylor} J.~E.,  {Navarro} J.~F.,  2001, ApJ, 563, 483

\bibitem[\protect\citeauthoryear{{Vikhlinin}, {Kravtsov}, {Forman}, {Jones},
  {Markevitch}, {Murray} \& {Van Speybroeck}}{{Vikhlinin}
  et~al.}{2006}]{Vikhlinin06}
{Vikhlinin} A.,  {Kravtsov} A.,  {Forman} W.,  {Jones} C.,  {Markevitch} M.,
  {Murray} S.~S.,    {Van Speybroeck} L.,  2006, ApJ, 640, 691

\bibitem[\protect\citeauthoryear{{Voit} \& {Donahue}}{{Voit} \&
  {Donahue}}{2005}]{Voit05}
{Voit} G.~M.,  {Donahue} M.,  2005, ApJ, 634, 955

\bibitem[\protect\citeauthoryear{{White} \& {Narayan}}{{White} \&
  {Narayan}}{1987}]{White87}
{White} S.~D.~M.,  {Narayan} R.,  1987, MNRAS, 229, 103

\end{thebibliography}

\end{document}